\documentclass[aps,prb,showpacs,twocolumn,floats,epsfig]{revtex4}
\usepackage{amssymb}
\usepackage{amsbsy}
\usepackage{amsmath}
\usepackage{epsfig}
\usepackage{textcomp}
\usepackage{graphicx}

\begin{document}

\title{Renormalization group approach to spinor Bose-Fermi mixtures
in a shallow optical lattice}

\author{S. Modak$^{1}$, S.-W. Tsai,$^{2}$ and K. Sengupta$^{1}$}

\affiliation{$^1$ Theoretical Physics Department, Indian Association
for the Cultivation of Science, Kolkata-700032, India. \\
$^2$ Department of Physics and Astronomy, University of California, Riverside,
California 92521, USA}

\date{\today}

\begin{abstract}

We study a mixture of ultracold spin-half fermionic and spin-one
bosonic atoms in a shallow optical lattice where the bosons are
coupled to the fermions via both density-density and spin-spin
interactions. We consider the parameter regime where the bosons are
in a superfluid ground state, integrate them out, and obtain an
effective action for the fermions. We carry out a renormalization
group analysis of this effective fermionic action at low
temperatures, show that the presence of the spinor bosons may lead
to a separation of Fermi surfaces of the spin-up and spin-down
fermions, and investigate the parameter range where this phenomenon
occurs. We also calculate the susceptibilities corresponding to the
possible  superfluid instabilities of the fermions and obtain their
possible broken-symmetry ground states at low temperatures and weak
interactions.

\end{abstract}

\maketitle

%\begin{center}

%\end{center}
\section{Introduction}
\label{intro}

The remarkable experimental achievements in the field of ultracold
atom physics have made it possible to generate mixtures of fermionic
atoms with different spin populations\cite{polarized_expmt}, as well
as mixtures of fermionic and bosonic atoms in a trap\cite{bf_expmt1}
that can also be loaded on optical lattices\cite{bf_expmt2}.
Multi-species fermions with unequal densities have also been
extensively studied not only in cold atom systems, but also in
electronic materials, such as the magnetic-field induced organic
superconductors\cite{organics} and other correlated fermion
systems\cite{electrons}, as well as in the context of color
superconductivity in dense quark matter\cite{quarks}. Bose-Fermi
mixtures present a rich phase diagram and have also been subject of
intense research\cite{bf_strong,bf_1d,bf_phasesep,
bf_staggered,blatter1,sk1,hops1,mediated1,mediated2,mediated3,mediated4}.
Several studies on such mixtures have been restricted to either
one-dimensional systems\cite{bf_1d} or to cases where the coupling
between the bosons and the fermions are weak
\cite{mediated1,mediated2,mediated3}. The existence of a supersolid
phase in these system in such a weak coupling regime has been
predicted\cite{blatter1}. Phase separation\cite{bf_phasesep} and
phases with staggered currents\cite{bf_staggered} have also been
investigated. Some of the other studies\cite{bf_strong}, which have
looked at the strong coupling regime, have restricted themselves to
integer filling factors of bosons and fermions or considered a
description of these systems at half filling either by using
analytical slave-boson mean-field technique\cite{sk1} or numerical
dynamical mean-field theory\cite{hops1}.

An interesting aspect of the study of quantum mixtures is that one
species of atoms may mediate interactions among atoms of the other
species. In a Bose-Fermi mixture where the bosons form a
Bose-Einstein condensate (BEC), quantum fluctuation of the BEC can
mediate long-range attractive interaction between the
fermions\cite{mediated1,mediated2,mediated3}. Conversely, in another
regime, fermions can be viewed as mediating an effective long-range
interaction between the bosonic atoms\cite{bf_phasesep}. In this
work we investigate the problem of partially polarized fermions
(unequal spin populations) in the presence of mediated interaction
due to quantum fluctuations of a BEC of bosonic atoms. Starting from
fermions with equal spin populations, we show that the spin
asymmetry of the fermion filling factors can arise due to coupling
to a spinor BEC. Spinor boson BEC systems have been studied both
experimentally\cite{spinor_expmt} and theoretically
\cite{spinor_theory,14}. In particular, the phases and low-energy
excitations of such a system is well-known. Here we consider the
effect of coupling of these excitations to the fermionic atoms in
the mixture.

%The presence of an optical lattice is well described by the Hubbard
%model, with on-site density-density interactions\cite{lattice}. In
%our study, however, for completeness, we consider a general
%Hamiltonian which contains both density-density and spin-spin
%interactions, and we calculate the effects of mediated interactions
%due the fluctuations of the spinor BEC.

An important tool for understanding the phases of interacting
fermions is the renormalization group (RG) technique\cite{shankar1}.
It has been applied to study the phase diagram of a Bose-Fermi
mixture with fermion interactions mediated by fluctuations of the
boson BEC, on square and triangular lattices\cite{mediated2}. The RG
for fermions has also been extended to frequency-dependent
interactions\cite{ret1}, where retardation effects are
important\cite{ret2,mediated3}. In this work, we use the
renormalization group technique to study a mixture of ultracold
spin-half fermionic and spin-one bosonic atoms in a shallow optical
lattice in two dimensions where the bosons are coupled to the
fermions via both density-density and spin-spin interactions. The
main aim of our study is to understand the effect of an
inter-species on-site SU(2) invariant spin-spin interaction on the
phases of this system. We consider the parameter regime where the
interaction between the bosons and the fermions are weak and the
bosons are in a superfluid state. We then start with a mean-field
treatment of the bosons, and include quantum fluctuations to first
order within a $1/N$ approximation. After a suitable Bogoliubov
transformation, the bosonic modes are integrated out, and an
effective action for the fermions is obtained. We find that, when
the bosons are in a spinor superfluid state, the spin-spin
interaction leads to an effective fermionic action with shifted
Fermi surfaces for the up- and down-spin fermions. We then carry out
a renormalization group analysis of this effective fermionic action
at low temperature and chart out the fate of such a shift under RG
flow for different parameter regimes. We also calculate the
susceptibilities corresponding to the superfluid instabilities of
the fermions and obtain the possible broken-symmetry fermionic
ground states at low temperature and weak interactions. In
particular we show that the leading instability for the fermions
with attractive interaction and circular Fermi surface occurs in the
triplet superfluid channel.

The organization of the rest of the paper is as follows. In Sec.\
\ref{seceff}, we introduce the model Hamiltonian for the Bose-Fermi
mixture and derive the effective fermionic action. In Sec.\
\ref{secrg1}, we obtain the RG equations for the fermionic self
energy and interactions from this action. Next, in Sec.\
\ref{secrg2}, we analyze the RG flow of different susceptibilities.
Finally, we present a discussion of our main results and conclude in
Sec.\ \ref{con}.

\section{Effective Fermionic Hamiltonian}
\label{seceff}

The Hamiltonian of a Bose-Fermi mixture in a shallow ultracold
lattice is given by $H =H_{F}+H_{B}+H_{BF}$. The fermionic part of
the Hamiltonian $H_F$ is given by
\begin{eqnarray}
 H_{F}&=&
\sum_{{\bf k} \sigma}(\varepsilon_{{\bf
k}}-\mu_{F})\widetilde{f}_{{\bf k}
\sigma}^{\dagger}\widetilde{f}_{{\bf k} \sigma}
\nonumber\\
&& +\sum_{{\bf q},{\bf k},{\bf k}^{\prime},\sigma}U^{1}_{{\bf
q}}{\widetilde{f}}^{\dagger} _{{\bf k}^{\prime}-{\bf q},\sigma}
\widetilde{f}_{{\bf k}^{\prime},\sigma}
\widetilde{f}^{\dagger}_{{\bf k}+{\bf q},\sigma}\widetilde{f}_{{\bf
k},\sigma}
\nonumber\\
&& +\sum_{{\bf q},{\bf k},{\bf k}^{\prime},\sigma}U^{2}_{{\bf
q}}{\widetilde{f}}^{\dagger} _{{\bf k}^{\prime}-{\bf q},\sigma}
\widetilde{f}_{{\bf k}^{\prime},\sigma}
\widetilde{f}^{\dagger}_{{\bf k}+{\bf q},{\bar
\sigma}}\widetilde{f}_{{\bf k},{\bar \sigma}},\label{i1}
\end{eqnarray}
where $\widetilde{f}_{{\bf k} \sigma}( \widetilde{f}_{{\bf k}
\sigma}^{\dagger})$ is the annihilation(creation) operator for the
fermions, $\mu_F$ denotes their bare chemical potential (taken to be
independent of the spin of the fermions), $U_{{\bf q}}^{1(2)}$
denotes the bare interaction between the fermions on the same
(separate) Fermi surfaces, $\varepsilon_{{\bf k}}= -2t_F [\cos(k_x
a) + \cos(k_y a)]$ is the fermion dispersion, $t_F$ is the hopping
amplitude of the fermions between the neighboring sites, ${\bar
\sigma}=\downarrow(\uparrow)$ for $\sigma=\uparrow(\downarrow)$, and
$a$ is the lattice spacing. For later use, we define the
two-component fermionic field $\phi_i= (\widetilde{f}_{i\uparrow},
\widetilde{f}_{i\downarrow})^T$ and use it to represent the
fermionic spin-density $S_{i
\gamma}^{F}=\phi_i^{\dagger}\sigma_{\gamma}\phi_i $ and number
density $n_i^F = \phi_i^{\dagger} \phi_i$, where $\vec \sigma =
(\sigma_x, \sigma_y, \sigma_z)$ denotes the Pauli matrices.

The Hamiltonian $H_B$ for the spinor bosons is given by \cite{14}
\begin{eqnarray}
H_{B}&=&-t_{b}\sum_{<i,j>,\alpha}\widetilde{b}_{i\alpha}^{\dagger}\widetilde{b}_{j\alpha}+\frac{U_{b0}}{2}
\sum_{i,\alpha}n_{i\alpha}^B (n_{i\alpha}^B -1)\nonumber\\
&+&\frac{U_{b2}}{2}\sum_{i,\alpha}(({\bf S}^B_{i})^2-2n_{i \alpha}^B)-\mu_{B}
\sum_{i,\alpha} n_{i \alpha}^B , \label{i2}
\end{eqnarray}
where $\alpha=-1,0,1$ denotes the azimuthal spin quantum number of
the bosons, $\widetilde{b}_{i\alpha}$ ($n_{i\alpha}^B=
\widetilde{b}_{i\alpha}^{\dagger} \widetilde{b}_{i \alpha}$) is the
bosonic annihilation (density) operator, $t_b$ is the boson hopping
amplitude between neighboring sites, $U_{b0}$ and $U_{b2}$ denote
the on-site boson interaction strengths in the spin-0 and spin-2
channels respectively, and $\mu_B$ is the chemical potential for the
bosons. The spin density of these bosons can be expressed in terms
of the generators of spin-one matrices: ${\bf S}_i^{B}
=\widetilde{b}_{i \alpha}^{\dagger}{\bf \lambda}_{\alpha
\beta}\widetilde{b}_{i \beta}$. The detailed expression for the
generators $ {\bf \lambda}$ is given in Appendix \ref{appa}.

The most general SU(2) invariant on-site interaction between the
bosons and the fermions is represented by $H_{BF}$. Note that since
the fermions carry spin half, conservation of azimuthal quantum
number $m_s$ does not preclude an on-site spin-spin interaction
between the bosons and the fermions. Thus we consider the
Hamiltonian $H_{BF}$ to be of the form
\begin{eqnarray}
H_{BF}&=& U_{ss}\sum_{i} {\bf S}_{i}^{F}\cdot {\bf
S}_{i}^{B}+U_{dd}\sum_{i}n_{i}^{F}n_{i}^{B}.\label{i3}
\end{eqnarray}
In what follows, we are going to consider the parameter regime
$U_{ss} \ll U_{dd} \ne 0$. We note that the presence of a non-zero
$U_{ss}$ is a key feature of the subsequent analysis carried out in
this work.

\begin{figure}
\includegraphics[width=0.6\linewidth]{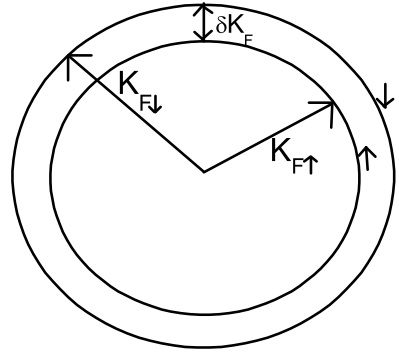}
\caption{Schematic representation of the separation between the up
and down spin Fermi surface. $\delta
K_F=(K_{F\downarrow}-K_{F\uparrow})$ is the difference between the
magnitudes of the up and the down spin Fermi momenta.} \label{fig6}
\end{figure}

The analysis of the coupled Bose-Fermi system is most easily done in
terms of coherent state path integrals. Following standard
prescription, we write the partition function of the system as

\begin{eqnarray}
Z &=& \int D[b] D[b^{\ast}] D[f] D[f^{\ast}]
e^{-S[f,f^{\ast},b,b^{\ast}]}, \nonumber\\
S &=& S_B + S_F + S_{BF}, \nonumber\\
S_{B} &=& - \frac{1}{\beta} \sum_{\Omega_n} \left[ \sum_{\bf k}
\left[ b^{\ast} ({\bf k},\Omega_n) i
\Omega_n b ({\bf k},\Omega_n)\right] - H_B[b^{\ast},b] \right], \nonumber\\
S_{F} &=& - \frac{1}{\beta} \sum_{\omega_n} \left[ \sum_{\bf k}
\left[ f^{\ast} ({\bf k},\omega_n) i
\omega_n f ({\bf k},\omega_n)\right] - H_F[f^{\ast},f] \right], \nonumber\\
S_{BF} &=& \frac{1}{\beta^2} \sum_{\omega_n,\Omega_n} H_{BF},
\label{ac1}
\end{eqnarray}
where $b = (b_1,b_0,b_{-1})$ [$f=(f_{\uparrow},f_{\downarrow})$]
denotes bosonic [fermionic] fields, $\Omega_n$ ($\omega_n$) denote
bosonic (fermionic) Matsubara frequencies, and $\beta=1/k_B T$ with
$T$ being the temperature, and $k_B$ the Boltzmann constant.

We begin with the analysis of $S_B$. We transform the Hamiltonian
written in Eq.\ (\ref{i2}) into $x,y,z$ basis by using the following
relations
\begin{eqnarray}
{b}_{x}^{*}&=&\frac{1}{\sqrt{2}}({b}_{-1}^{*}-{b}_1^{*}) \ , \nonumber\\
{b}_{y}^{*}&=&\frac{i}{\sqrt{2}}({b}_{-1}^{*}+{b}_1^{*}) \ , \nonumber\\
{b}_z^{*}&=&{b}_0^{*}.\label{basis}
\end{eqnarray}
We assume that the bosonic spinor system is deep in the BEC state.
The standard procedure for analyzing such a BEC involves expressing
the bosonic field $b$ as
\begin{eqnarray}
{b}_{\alpha}({\bf q},i\Omega_n) &=&\Psi_{0\alpha}\delta_{{\bf
q},0}+{a}_{\alpha} ({\bf q},i\Omega_n), \label{i5}
\end{eqnarray}
where $ \Psi_{0\alpha} = \langle{b}_{\alpha}({\bf q}=0,\Omega_n=0)
\rangle$, and expanding $S_B$ to order $O(a_{\alpha}^2)$. The
mean-field equation for the condensate is then obtained by imposing
the coefficient of $a_0$ and $a_0^{\dagger}$ to be zero. This
analysis yields
\begin{eqnarray}
\Psi_{0\alpha}^{*} \left[ - 4t_{b} + \left(n_{0} - \frac{1}{2} \right) U_{b0}
+ (n_{0} - 1) U_{b2} - \mu_{b} \right]\nonumber\\
- \frac{U_{b2}}{N} \sum_{\beta \ne \alpha}
(\Psi_{0\beta}^{*})^{2}\Psi_{0\alpha} = 0,\label{i8}
\end{eqnarray}
where we have introduced the condensate density $n_{0}=N_{b}/N $,
with $N_{b}$ being the total number of bosons in the system, and $N$
the total number of lattice sites. As shown in a previous study,
Eq.\ (\ref{i8}) supports two solutions\cite{14}. The first is a
ferromagnetic phase with
\begin{eqnarray}
\Psi_{\rm ferro}=\sqrt{n_{0}N/2}(1,\pm i,0)^{T}, \label{ferrosol}
\end{eqnarray}
while the other is a polar phase with $\Psi_{\rm polar} =
\sqrt{n_{0}N} (1,0,0)^{T}$. The ferromagnetic state becomes
energetically favorable for $U_{b2}<0$ and in the rest of the paper
we concentrate on this regime and work with the solution $\Psi_{\rm
ferro}=\sqrt{n_{0}N/2}(1,i,0)^{T}$ which corresponds to $\langle b_1
\rangle \ne 0$. The choice of one of the these two solutions can be
easily seen to be the effect of any stray magnetic field that might
be present in a realistic experimental system. We note that for the
polar phase, the low-energy physics of the Bose-Fermi mixture is
identical to its counterpart with spinless bosons
\cite{mediated2,mediated3}.

A detailed analysis of $S_{B}$ when the bosons are in the
ferromagnetic phase is carried out in Appendix \ref{appa} and leads
to the expression for $S_B$ [Eq.\ (\ref{a2})] which is quadratic in
the fluctuation fields $a_{\alpha}({\bf q},i\Omega_n)$. Thus, using
Eq.\ (\ref{a2}) and Eq.\ (\ref{i3}), one can integrate out the boson
degrees of freedom and obtain, after a straightforward but tedious
calculation, an effective action for the fermions,
\begin{eqnarray}
S_{\rm eff} &=& - \frac{1}{\beta} \sum_{i\omega_n}  \Big[ \sum_{{\bf
k}}f^{\ast}_{{\bf k},\sigma}[i \omega_n - \varepsilon_{{\bf
k}} + \mu_{F \sigma}] f_{{\bf k},\sigma}\nonumber\\
&& -\sum_{{\bf k,k^{\prime},q},\sigma} \widetilde{U}_{{\bf
q},\sigma\sigma}f^{\ast}_ {{\bf k}+{\bf q},\sigma}f_{{\bf k},\sigma}
f^{\ast}_{{\bf k'-q},\sigma}f_{{\bf k'},\sigma}
\nonumber\\
&& - \sum_{{\bf k,k^{\prime},q}} \widetilde{U}_{{\bf
q},\uparrow\downarrow}f^{\ast}_{{\bf k+q},\uparrow}f_{{\bf
k},\downarrow} f^{\ast}_{{\bf k'-q},\downarrow}f_{{\bf k'},\uparrow}
\Big] \ , \label{fham1}
% \nonumber\\
% && + \sum_{{\bf k,k^{\prime},q},\sigma\neq\overline{\sigma}}
% \widetilde{U}^3_{{\bf q}} f^{\dagger}_{{\bf
% k^{\prime}-q},\sigma}f_{{\bf k}^{\prime},
% \overline{\sigma}}f^{\dagger}_{{\bf k+q},\overline{\sigma}}f_{{\bf
% k},\sigma}
\end{eqnarray}
where $\mu_{F \sigma}= \mu_{F}-[n_{0}U_{ss}({\rm Sgn}\sigma)+
U_{dd}n_{0}]$, and ${\rm Sgn} \sigma =1(-1)$ for
$\sigma=\uparrow(\downarrow)$. Thus the spin-spin interaction
between the bosons and the fermions leads to an effective shift
between the spin-up and spin-down Fermi surfaces, as shown in Fig.\
\ref{fig6} , at the mean-field level. The sign of the shift depends
on the choice of one of the two solutions given by Eq.\
\ref{ferrosol}; for our choice $\Psi_{\rm
ferro}=\sqrt{n_{0}N/2}(1,i,0)^{T}$, the down-spin Fermi surface is
enhanced compared to the up-spin one as shown in Fig.\ \ref{fig6}.
The effective interactions $\widetilde{U}_{{\bf q},\sigma\sigma}$
and $\widetilde{U}_{{\bf q},\uparrow\downarrow}$ are given by
\begin{eqnarray}
\widetilde{U}_{{\bf q},\sigma\sigma} &=& U_{\bf
q}^1-\frac{n_0}{2}(U_{dd}+({\rm Sgn}\sigma)
U_{ss})^2\chi_{{\bf q},\sigma\sigma}, \label{ii13}\\
\widetilde{U}_{{\bf q},\uparrow\downarrow} &=& U_{\bf q}^2+
\frac{n_0}{2}(U_{dd}^2-U_{ss}^2)\chi_{{\bf q},\uparrow\downarrow},
\label{i13}
\end{eqnarray}
where
\begin{eqnarray}
\chi_{q,\sigma\sigma'}=\frac{\xi_{{\bf q},\sigma\sigma'}-2U_{b2}n_0}
{(\xi_{{\bf q},\sigma\sigma'}-2U_{b2}n_0)^2+\Omega_n^2},\label{chi}
\end{eqnarray}
and $\xi_{{\bf q},\sigma\sigma'}=-2t_b[\cos(q_{x
\sigma\sigma'})+\cos(q_{y \sigma\sigma'})-2]$ is the boson
dispersion at the wave-vector $q_{x[y] \sigma \sigma} = K_{F \sigma}
\cos(\theta) [\sin(\theta)]$, $q_{x[y] \sigma {\bar \sigma}} = (K_{F
\sigma}+ K_{F {\bar \sigma}}) \cos(\theta) [ \sin (\theta) ]/2$.
Here we have set the lattice spacing $a=1$, $K_{F\sigma}$ is the
magnitude of the Fermi wave-vector for electrons with spin $\sigma$,
and we have restricted ourselves to the regime where $|K_{F
\uparrow}-K_{F \downarrow}| \ll K_{F \uparrow},K_{F \downarrow}$.
This restricts the validity of our analysis to the parameter regime
$\mu_F \gg U_{ss}$. Note that ${\widetilde U}_{{\bf
q},\uparrow\downarrow}$ represents the amplitude of scattering
between fermions on separate Fermi surfaces, while scattering
processes represented by ${\widetilde U}_{{\bf q},\sigma\sigma}$
involve fermions on the same Fermi surface. For the rest of the
paper, we shall ignore the retardation effects of the effective
interaction and shall thus set $\Omega_n=0$ and restrict ourselves
to circular Fermi surfaces with small effective shifts between them
(as shown in Fig.\ \ref{fig6}).

Next, following standard procedure outlined in Ref.\
\onlinecite{shankar1}, we antisymmetrize $\widetilde{U}_{{\bf
q},\sigma\sigma}$ with respect to the interchange ${\bf
k_1}\leftrightarrow {\bf k_2}$ and ${\bf k_3}\leftrightarrow {\bf
k_4}$ (where, ${\bf k_1} ={\bf k'},~{\bf k_2} = {\bf k} ,~ {\bf k_3}
= {\bf k'-q},~ {\bf k_4 }= {\bf k + q}$). Further, following Ref.\
\onlinecite{shankar1} and using the circular nature of the Fermi
surface, we consider fermion scattering only in the forward (${\bf
k_1}={\bf k_4 }$ and $ {\bf k_2}={\bf k_3})$ and BCS (${\bf
k_2}=-{\bf k_1} $ and $ {\bf k_4}=-{\bf k_3}$) channels. The
contribution of $\widetilde{U}_{{\bf q}, \sigma \sigma}$ to these
channels can be computed from Eqs.\ (\ref{i13}) and (\ref{chi}).
Denoting interaction couplings in these channels by
$\widetilde{F}_{\sigma\sigma}(\theta_{12})$ and
$\widetilde{V}_{\sigma\sigma}(\theta_{13})$, respectively, we find
that
\begin{eqnarray}
\widetilde{F}_{\sigma\sigma}(\theta_{12}) &=&
\left[U_0-n_0\left(\frac{U_{dd}+({\rm Sgn} \sigma)
U_{ss}}{2n_0U_{b2}}\right)^2K_{F\sigma}^2\right] \nonumber\\
&& \times [1- \cos(\theta_{12})], \label{antsym1} \\
\widetilde{V}_{\sigma\sigma}(\theta_{13}) &=&
\left[U_0-n_0\left(\frac{U_{dd}+({\rm Sgn} \sigma)
U_{ss}}{2n_0U_{b2}}\right)^2K_{F\sigma}^2\right]
\nonumber\\
&& \times \cos(\theta_{13}), \label{antsym2}
\end{eqnarray}
where, $\theta_{12}(\theta_{13})$ is the angle between ${\bf k_1}$
and ${\bf k_2 }({\bf k}_3)$, $U_0
(1-\cos(\theta_{12}))[U_0\cos(\theta_{13})]$ denote the value of
$U_{\bf q}^{1}$ for the forward[BCS] channels. Note that in
obtaining Eq.\ (\ref{antsym1}) and (\ref{antsym2}), we have
explicitly antisymmetrized the contribution of $\chi_{{\bf q},
\sigma \sigma}$ in Eq.\ (\ref{ii13}).

The  contributions of $\widetilde{U}_{{\bf q},\uparrow\downarrow}$ in the
forward and the BCS channels can also be computed in a similar manner
and are given by
\begin{eqnarray}
\widetilde{F}_{\uparrow\downarrow}(\theta_{12})&=&
\frac{n_0(U_{dd}^2-U^2_{ss})}{4K^2_{F,\uparrow\downarrow}
(1-{\cos}(\theta_{12}))-4n_0U_{b2}},\nonumber\\
\widetilde{V}_{\uparrow\downarrow}(\theta_{13}) &=&
\frac{n_0(U_{dd}^2-U^2_{ss})}{4K^2_{F,\uparrow\downarrow}
\cos(\theta_{13})-4n_0U_{b2}},\label{ud1}
\end{eqnarray}
where, $K_{F,\uparrow\downarrow} =(K_{F\uparrow}+K_{F\downarrow})/2$
and we have set the contribution of $U_{{\bf q}}^2$ to the forward
and BCS channels to zero. We have checked explicitly that finite
value of $U_{\bf q}^2$ does not alter the qualitative conclusions of
the work.

\section{RG equations for self-energy and couplings}
\label{secrg1}

In this section, we carry out a RG analysis of $S_{\rm eff}$
adapting one-loop Wilsonian RG using a path integral approach. The
details of this approach are outlined in several past works
\cite{shankar1,ret1,ret2}. The key idea behind such a procedure is
to consider $S_{\rm eff}$ as the starting fermionic action at a
high-energy cutoff scale $\Lambda$, perform Wilson RG on this action
and derive the flow equations for the effective interactions and
fermionic self-energy. It is well known \cite{shankar1} that for a
circular Fermi surface as considered here, only the interaction in
the BCS channels ($V$) flow under RG and that the key contribution
to the fermionic self-energy within one-loop RG comes from the
forward channels ($F$).

\begin{figure}
\includegraphics[width=\linewidth]{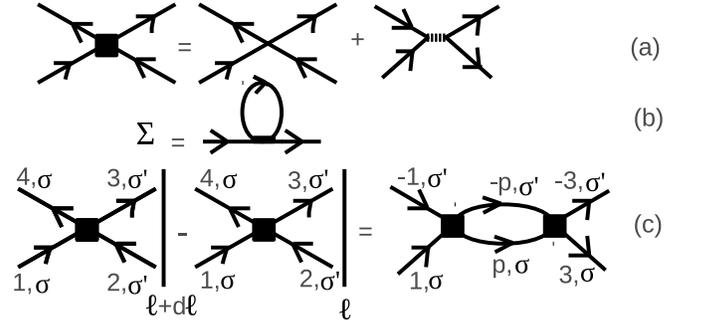}
\caption{(a) Schematic representation of the effective fermionic
interaction $\widetilde U$ (left) as sum of the bare interaction
$U_0$ and the contribution from the bosons. (b) Diagrammatic
representation of the RG flow equation for the self-energy
($\Sigma_{\sigma}$) correction which receives contribution from the
interaction in the forward channel $F$. (c) Diagrammatic
representation for the RG flow equations for $V$. We use the
simplified notation $1={\bf k^{\prime}}$, 2=${\bf k}$,
$3={\bf k}'-{\bf q}$, $4={\bf k}+{\bf q}$, and $\sigma$ and $\sigma^{\prime}$
takes values $\uparrow$ and $\downarrow$. }\label{fig7}
\end{figure}

\begin{figure}
\includegraphics[width =\linewidth]{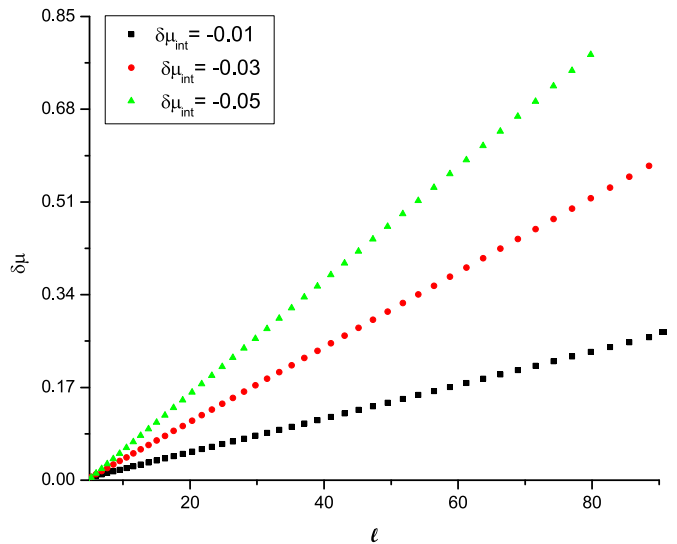}
\caption{Plot of $\delta\mu~{\rm versus}~\ell$, for $ l=0$,
$n_{0}=0.9$, $U_{b2}=-0.08$, $U_0=-0.3$ for several values of
$\delta \mu_{\rm int}=-0.05$ (green line with triangles),
$\delta\mu_{int}=-0.03$ (red line with circles) and
$\delta\mu_{int}=-0.01$ (black line with squares).} \label{fig2}
\end{figure}

\begin{figure}
\includegraphics[width =\linewidth]{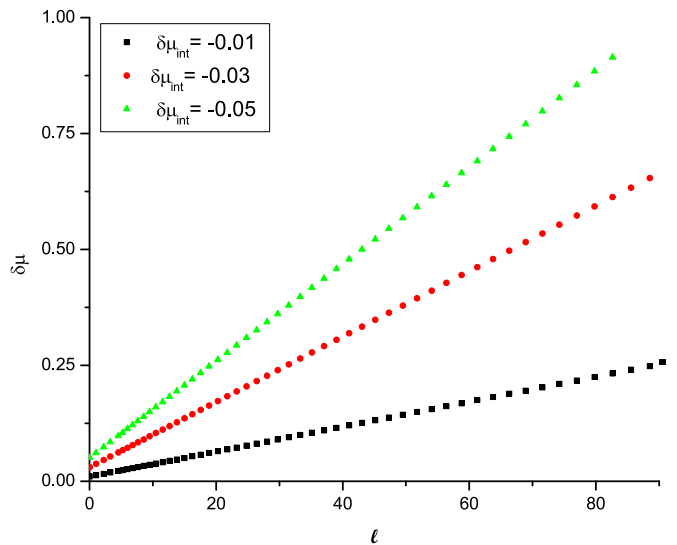}
\caption{Same as in Fig.\ \ref{fig2} but with $U_0=$0.3.}
\label{fig2a}
\end{figure}

Using these facts, the relevant diagrams for the contribution to the
self-energy and the effective interactions in the present model
[Eq.\ (\ref{fham1})] can be easily found. These are shown in Fig.\
\ref{fig7}. The RG equations for the interactions and the fermionic
self-energy, as obtained from these diagram in Fig.\ \ref{fig7}, are
given by,
\begin{eqnarray}
\frac{d\Sigma_{\sigma}(\theta)}{d\ell}&=&-\frac{1}{2\pi}\int_{\theta'
\omega_n} \Big(K_{F \sigma}
\widetilde{F}_{\sigma\sigma}(\theta^{\prime}-\theta)G_{\sigma}(\omega_n,\theta')
\nonumber\\
&& + K_{F {\bar \sigma}}\widetilde {F}_{\uparrow\downarrow}
(\theta^{\prime}-\theta) G_{\bar{\sigma}}(\omega_n,\theta')\Big),
\label{i115} \\
\frac{d\widetilde{V}_{\sigma \sigma }(\theta_1-\theta_3)}{d\ell}&=&
-\frac{K_{F \sigma}}{2\pi} \int_{\theta \omega_n}
\widetilde{V}_{\sigma \sigma}
(\theta_1-\theta)\widetilde{V}_{\sigma \sigma}(\theta-\theta_3)\nonumber\\
&& \times G_{\sigma}(\omega_n,\theta) G_{\sigma}(-\omega_n,
\theta+\pi),\label{i16} \\
\frac{d\widetilde{V}_{\uparrow\downarrow}(\theta_1-\theta_3)}{d\ell}&=&-\frac{K_{F
\uparrow\downarrow}} {2\pi}\int_{\theta \omega_n}
\widetilde{V}_{\uparrow\downarrow}
(\theta_1-\theta)\widetilde{V}_{\uparrow\downarrow}
(\theta-\theta_3)\nonumber\\
&& \times G_{\downarrow}(\omega_n,\theta) G_{\uparrow}(-\omega_n,
\theta+\pi), \label{i18}
\end{eqnarray}
where we have carried out the integrals over the radial momentum
perpendicular to the circular Fermi surface,
$\Lambda=\Lambda_{0}e^{-\ell}$ is the RG cutoff, $\Lambda_{0}<
E_{F}$ is the cut-off in the beginning of the RG flow, ${\ell}$ is
the RG time, $\int_{\theta' \omega_n} = 1/\beta \sum_{i \omega_n}
\int d\theta'/(2\pi)$ denotes frequency sum and integral over
transverse momenta over the Fermi surface, $\Sigma_{\sigma}(\theta)$
denotes the self-energy for fermions with spin $\sigma$ and momentum
${\bf k}=(K_{F\sigma} \cos(\theta),K_{F\sigma} \sin(\theta))$,
$K_{F,\sigma\overline{\sigma}} =(K_{F\uparrow}+K_{F\downarrow})/2$,
$\epsilon_{\sigma}(\theta')$ is the fermion dispersion on the Fermi
surface with spin $\sigma$ , and the fermion Green function,
evaluated on the Fermi surface for spin $\sigma$ electrons are given
by
\begin{eqnarray}
G_{\sigma}(\omega_n,\theta) = \left(i\omega_n
-(\epsilon_{\sigma}(\theta)-\mu_{\sigma})
-\Sigma_{\sigma}(\theta)\right)^{-1}. \label{fgf}
\end{eqnarray}
%carry out the Matsubara sum over $\omega_n$
Before solving Eqs.\ (\ref{i115})..(\ref{i18}) numerically, we note
that $\epsilon_{\sigma}(\theta)= -2t [\cos(K_{F\sigma} \cos(\theta))
+\cos(K_{F \sigma} \sin(\theta))]$ have a very weak $\theta$
dependence. Further, the integration
$\widetilde{F}_{\sigma\sigma'}(\theta'-\theta)$ over $\theta'$ for a
complete cycle renders it independent of $\theta$ as well.
Consequently, $\Sigma_{\sigma}$ becomes independent of $\theta$.
Thus, at low temperature, $\sum_{\omega_n}
G_{\sigma}(\omega_n,\theta) G_{\sigma'}(-\omega_n, \theta+\pi)$
becomes practically independent of $\theta$. Using this fact, it is
possible to express Eqs. (\ref{i115})..(\ref{i18}) in the angular
momentum channels denoted by $l$ to obtain
\begin{eqnarray}
% \frac{d\Sigma_{\sigma}^l}{d\Lambda}&=&-\frac{1}{2\pi}\Big(K_{F \sigma}
% \widetilde{F}_{\sigma\sigma}^l\frac{1}{\beta}\sum_{\omega_n}G_{\sigma}(\omega_n)\nonumber\\
% &+& K_{F {\bar \sigma}}
% \widetilde {F}_{\uparrow\downarrow}^l\frac{1}{\beta}\sum_{\omega_n}G_{\bar{\sigma}}(\omega_n)\Big)\label{l1} \\
\frac{d\Sigma_{\sigma}^l}{d\ell}&=& -\frac{1}{2\pi}\Big(K_{F
\sigma} \widetilde{F}_{\sigma\sigma}^lG_{\sigma}^l+ K_{F {\bar
\sigma}}
\widetilde {F}_{\uparrow\downarrow}^l G_{\bar{\sigma}}^l\Big),\label{l1} \\
&& \frac{d\widetilde{V}_{\sigma \sigma }^l}{d\ell}\simeq
-\frac{K_{F
\sigma}J_{\sigma\sigma}}{2\pi} (\widetilde{V}_{\sigma \sigma} ^l)^2, \\
&& \frac{d\widetilde{V}_{\uparrow\downarrow}^l}{d\ell}
\simeq-\frac{K_{F \uparrow\downarrow}J_{\uparrow \downarrow}}{2\pi}
(\widetilde{V}_{\uparrow\downarrow}^l)^2. \label{l4}
\end{eqnarray}
where
$\widetilde{V}_{\sigma\sigma'}^l[\widetilde{F}_{\sigma\sigma'}^l]$ and $G^l_{\sigma}$
are given by
\begin{eqnarray}
%G_{\sigma}^l&=&\sum_{\theta}e^{il\theta}(1/\beta\sum_{i\omega_{n}} G_\sigma(\omega_n,\theta))\label{gl}\\
\widetilde{V}_{\sigma\sigma'}^l[\widetilde{F}_{\sigma\sigma'}^l]&=&\int_0^{2
\pi} \frac{d \theta}{2\pi}
e^{il\theta}\widetilde{V}_{\sigma\sigma'}(\theta)
[\widetilde{F}_{\sigma\sigma'}^l(\theta)],\label{h}\\
G_{\sigma}^l&=& \int_0^{2 \pi} \frac{d \theta}{2\pi}
e^{il\theta}(1/\beta\sum_{i\omega_{n}}
G_\sigma(\omega_n,\theta))\label{gl}, \nonumber\\
J_{\sigma \sigma} &=& \frac{1}{\beta} \sum_{i \omega_n} G_{\sigma}
(\omega_n, \theta) G_{\sigma} (-\omega_n, \theta+\pi),  \nonumber\\
J_{\uparrow \downarrow} &=& \frac{1}{\beta} \sum_{i \omega_n}
G_{\uparrow} (\omega_n, \theta) G_{\downarrow} (-\omega_n,
\theta+\pi).
\end{eqnarray}
Next, we solve the RG equations for self energy and couplings
numerically for a temperature $\beta t_F=10$. We have carried out
the numerical solution of both Eqs.\ (\ref{i115})..(\ref{i18}) and
Eqs.\ (\ref{l1})..(\ref{l4}) and checked that these yield identical
results confirming our observation on the absence of $\theta$
dependence of $J_{\sigma\sigma}$ and $J_{\uparrow \downarrow}$. For
the numerical solution of these equations, we have scaled all the
energy parameters in units of $2t_{F}$. We note that at the one-loop
level, the effect of $\Sigma_{\sigma}$ is to renormalize the
chemical potential $\mu_{F\sigma}$ and hence their difference:
$\delta\mu=|\mu_{F\uparrow}-\mu_{F\downarrow}+\Sigma^l_{\uparrow}
-\Sigma^l_{\downarrow}|= |\delta \mu_{int} +\Sigma^l_{\uparrow}
-\Sigma^l_{\downarrow}|$, where $\delta \mu_{int}=\mu_{F\uparrow}-
\mu_{F\downarrow}=-2n_0U_{ss}$. In Figs.\ \ref{fig2} and
\ref{fig2a}, we show the variation of $\delta\mu$ as a function of
the RG time $\ell$ for $U_{0}=\pm 0.3$, $n_0=0.9$, $U_{b2}=-0.08$
and three different representative values of $\delta\mu_{int}$. We
find that the RG flow is essentially controlled by the induced
interaction part of $F_{\sigma\sigma'}$ and displays little
dependence on $U_0$. The separation between the Fermi surfaces is
always amplified and the spin-up and spin-down Fermi surfaces flow
away from each other. This signifies a possibility of either a
Ferromagnetic or triplet superfluid (with equal-spin pairing)
instabilities of the system. Note that such instabilities, in case
they occur, have their root in the initial separation of the
opposite spin Fermi surfaces and hence can be attributed to the
spin-spin coupling between the fermions and the spinor bosons.
\begin{figure}
 \includegraphics[width=\linewidth]{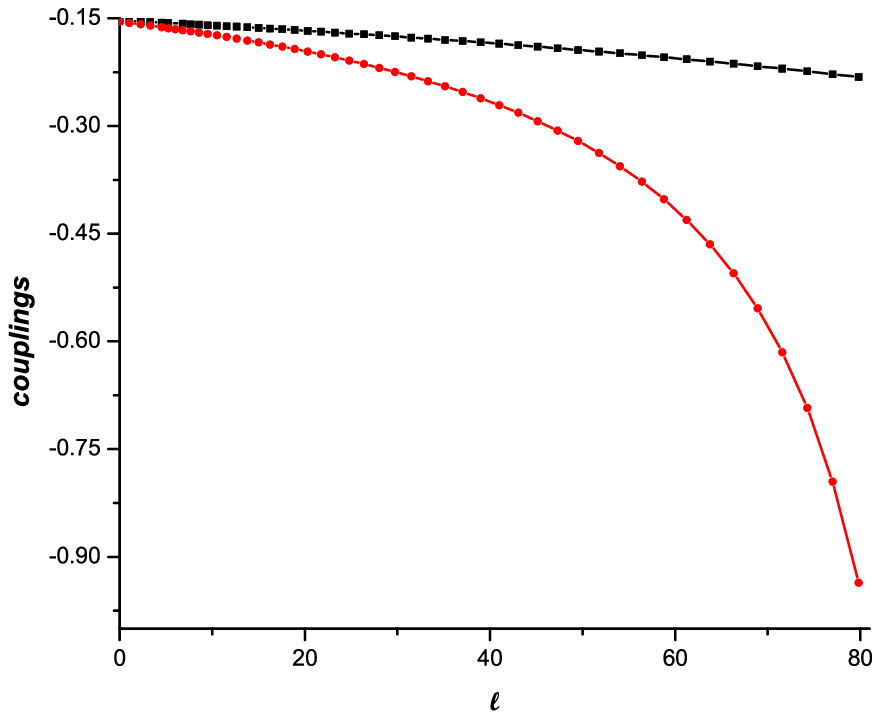}
\caption{RG flow of the couplings $\widetilde{V}_{\uparrow\uparrow}$
(black line with squares), $\widetilde{V}_{\downarrow\downarrow}$
(red line with circles) for $l=\pm 1$, $U_0$= -0.3, $U_{b2}=-0.08$,
$\delta \mu_{int}=-0.05$, and $n_0=0.9$.} \label{fig3}
\end{figure}
\begin{figure}[thb] \includegraphics[width=\linewidth]{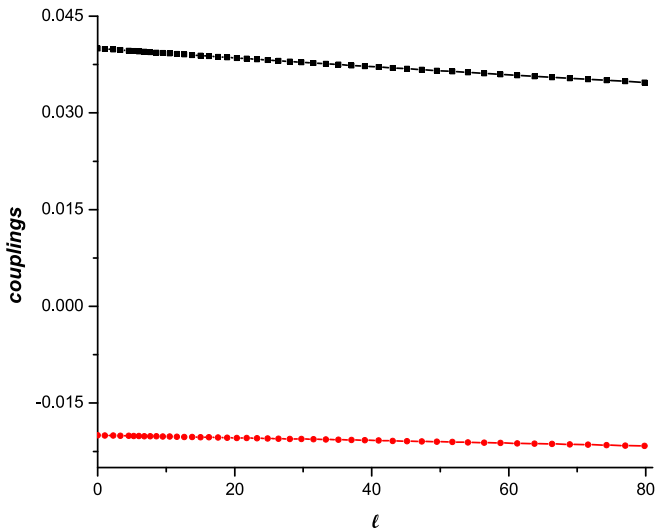}
\caption{RG flow of the coupling
$\widetilde{V}_{\uparrow\downarrow}$ (black line with squares for
l=0), $\widetilde{V}_{\uparrow\downarrow}$ (red line with circles
for $l=\pm 1$) for  $U_0$= -0.3, $U_{b2}=-0.08$, $\delta
\mu_{int}=-0.05$, and $n_0=0.9$.} \label{fig4}
\end{figure}

Next we plot the variation of the couplings with the RG cutoff
$\ell$ in Fig.\ \ref{fig3} and Fig.\ \ref{fig4} for $\delta
\mu_{int}=-0.05$, $n_0=0.9$, $U_{b2}=-0.08$ and  $U_{0}=- 0.3$. We
find that $\widetilde{V}_{\uparrow\downarrow}$ does not flow
appreciably under RG which is a consequence of lack of scattering
between Fermi surfaces with opposite spins. The flow of
$\widetilde{V}_{\sigma \sigma}$ shows an increase of their magnitude
indicating a flow toward strong coupling regime which can not be
accessed by our perturbative RG analysis.

\section{RG flow of the susceptibilities}

\label{secrg2}

In this section, we consider the RG flow for the possible
instabilities of the fermionic models. In particular, we consider
the singlet and equal-spin paired triplet superfluid (SSF and TSF)
instabilities \cite{8} of the metallic phase of the fermions due to
the induced interaction. This choice is motivated by the fact that
for circular Fermi surfaces considered here we do not have nesting
and hence do not expect to have instabilities in the $2k_F$ spin- or
charge-density wave channels. It is well known that the onset of
such instabilities are signalled by the divergence of the
corresponding static susceptibilities under RG flow\cite{17}.

The flow equations for the static susceptibilities can be derived
using standard techniques as elaborated in Refs.\
\onlinecite{17,ret2,mediated3}. As outlined in these works, the
response function can be calculated by introducing a source term in
the action
\begin{eqnarray}
S_h=-\sum_q h^{\delta} \Delta^{\delta}, \label{source terrm}
\end{eqnarray}
where $\delta$ takes values SSF or TSF corresponding to the singlet
or equal-spin triplet pairings.
\begin{eqnarray}
\Delta^{{\rm SSF}}&=&\sum_{\sigma, {\bf
k}}(\rm{sgn}\sigma) f_{\sigma}({\bf k})f_{\bar{\sigma}}({\bf -k}) ~~~{\rm and}\nonumber\\
\Delta^{{\rm TSF}}&=& f_{\sigma}({\bf k})f_{\sigma}({\bf -k})
\label{ordpara},
\end{eqnarray}
are the order parameters for singlet and triplet superfluidity
respectively and $h^{\delta}$ is the external field of type
$\delta$. The corresponding response function is given by
\begin{eqnarray}
\chi^{ \delta} &=&\langle {\Delta^{ \delta}}^* \Delta^{ \delta}
\rangle = \frac{\delta^{(2)}\rm{ln}Z[h]}{\delta {h^{\delta}}^*
\delta h^{\delta}}\Big|_{h=0}\label{chisource},
\end{eqnarray}
where $Z[h]$ denotes the partition function in presence of the
source term. The RG process generates correction to the source field
$h$ along with the higher order terms in the source field. At any RG
time $\ell$, the total action $S_{\ell}$ can be written as
 \begin{eqnarray}
S_{\ell}&=& S_{\ell}^0 -\int dt \, \left[ z^{\delta} h^{\delta}
\Delta^{\delta} - {h^{\delta}}^* h({\bf q}) \chi^{\delta} \right]
,\label{z}
 \end{eqnarray}
where $S^0_{\ell}$ is the action at RG time $\ell$ without the
external field $h$ and the coefficient $z^{\delta}$ is the effective
vertex of type $\delta$.

\begin{figure}
\includegraphics[width=\linewidth]{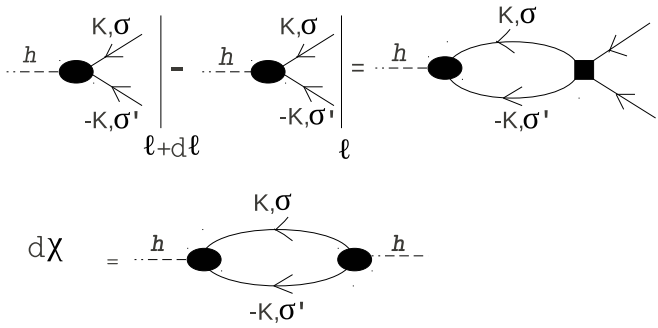}
\caption{Diagrammatic representation of the renormalization of
vertices and the susceptibilities in the BCS channel.} \label{fig8}

\end{figure}

\begin{figure} \includegraphics[width=\linewidth]{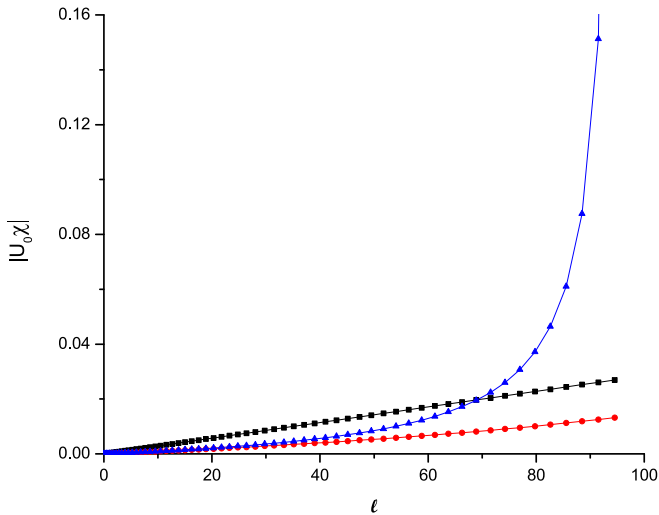}
\caption{RG flow for the static susceptibilities for $U_0=-0.3$. The
blue line with triangle and the black line with square indicate RG
flow of the susceptibilities for the triplet BCS channels
$\chi_{l,\downarrow\downarrow}^{{\rm TSF}}~(l=\pm 1)$ and
$\chi_{l,\uparrow \uparrow}^{{\rm TSF}}~(l=\pm 1)$ respectively. The
red line with circle indicates RG flow of the susceptibility for the
singlet BCS channel $\chi_{l}^{{\rm SSF}}~(l= 0)$.} \label{fig5}
\end{figure}

The relevant one-loop diagrams representing the RG equations for the
vertices $z^{\delta}$ and the susceptibilities $\chi^{\delta}$
are schematically shown in Fig.\ \ref{fig8}. The corresponding
one-loop flow equations for $z^{\delta}$ and $\chi^{\delta}$ are
given by

\begin{eqnarray}
\frac{dz^{\rm{TSF}}_{l,\sigma\sigma}}{d\ell}&=&-\frac{K_{F\sigma}}{2\pi}J_{\sigma\sigma}
\widetilde{V}_{\sigma\sigma}^lz^{\rm{TSF}}_{l,\sigma\sigma}, \nonumber\\
\frac{d\chi^{\rm{TSF}}_{l,\sigma\sigma}}{d\ell}&=&\frac{K_{F\sigma}}{2\pi}J_{\sigma\sigma}(z^{\rm{TSF}}_{l,\sigma\sigma})^2,
\nonumber\\
\frac{dz^{\rm{SSF}}_{l}}{d\ell}&=&-z^{\rm{SSF}}_l\sum_{\sigma}\frac{K_{F,\sigma\bar\sigma}}{2\pi}({\rm
sgn}\sigma)
J_{\sigma\bar\sigma} \widetilde{V}_{\sigma\bar\sigma}^l, \nonumber\\
\frac{d\chi^{\rm{SSF}}_{l}}{d\ell}&=&(z^{\rm{SSF}}_l)^2\sum_{\sigma}\frac{K_{F,\sigma\bar\sigma}}{2\pi}J_{\sigma\bar\sigma},
\label{i26}
\end{eqnarray}
where we have used the $\theta$ independence of $J_{\sigma\sigma}$
and $J_{\uparrow\downarrow}$. We solve these equations numerically
for $\beta t_F=10 $. The results are shown in Fig.\ \ref{fig5} for
$U_0=-0.3$ for SSF and TSF instabilities. We find that the $\chi_{l,
\downarrow\downarrow}^{{\rm TSF}}$ instability shows a divergence
around $\ell \simeq 80$ indicating an instability of the metallic
ground states against triplet down-spin pairing superfluid ground
state. This is an expected consequence of the growing separation of
the Fermi surfaces which prevents opposite spin SSF pairing and
hence favors down-spin TSF state. Thus we conclude that the most
dominant instability of the Fermi superfluid with an attractive
interaction is TSF with down-spin pairing. We note that our RG
analysis can not predict the subsequent fate of the system once the
superfluid instability has set in. The system may either end up with
a spin-up metallic Fermi surface coexisting with a triplet
superfluid of spin-down fermions or the superfluidity in the
spin-down channel may induce a superfluid instability for the
spin-up fermions via a momentum-space proximity effect. The latter
effect is somewhat similar to that seen for multi-band ruthenate
superconductors \cite{rice1}. We leave a more thorough analysis of
these possibilities as a subject of future study.


\section{Conclusion}
\label{con} In conclusion, we have studied a mixture of spinor boson
and fermion in a shallow 2D optical lattice using RG and have shown
that the presence of an on-site spin-spin interaction between the
bosons and the fermions leads to a separation of Fermi surface of
the spin-up and spin-down fermions irrespective of the nature of the
bare interaction between the fermions provided that the bosons are
in the spinor condensate state. Such a separation, depending on the
density of the fermions, may give rise to a net spin polarization
for the Fermi superfluid. Further, for attractive interaction
between these fermions, we have shown that the leading instability
of the metallic state of the fermions lies in the TSF channel with
down spin pairing. In particular, we predict that for fermions
coupled to a spinor bosonic condensate in its ferromagnetic phase
via a spin-spin interactions, attractive interactions will induce a
down-spin triplet pairing superfluid instability over the otherwise
more common singlet pairing instability. We note that this
phenomenon is in contrast to the fermions coupled to either a
spinless boson condensate or a spinor boson condensate in its polar
phase.

\section{Acknowledgements}

It is our pleasure to thank Filippos Klironomos for invaluable
discussions. SWT gratefully acknowledges support from NSF under
grant DMR-0847801 and from the UC-Lab FRP under award number
09-LR-05-118602. KS thanks DST, India for support through grant
SR/S2/CMP-001/2009.

\appendix

\section {Effective Quadratic Hamiltonian for spinor boson}
\label{appa}

The generators $\lambda_+$, $\lambda_-$ for spin-one bosons can be
obtained from the spin-rotation matrices in $x$,$y$ and $z$ basis.
In this basis we have

$ S_x=\frac{1}{\sqrt{2}}\left(
\begin{array}{ccc}
0& 1 & 0 \\
1& 0 & 1 \\
0 & 1 & 0
\end{array}
\right)$ ,
$S_y=\frac{1}{\sqrt{2}}\left(
\begin{array}{ccc}
0& -i & 0 \\
i& 0 & -i \\
0 & i & 0
\end{array}
\right)$ ,

 and  $ S_z=\left(
\begin{array}{ccc}
1& 0 & 0 \\
0& 0 & 0 \\
0 & 0 & -1
\end{array}
\right)$.

This yields, using $\lambda_{\pm} = S_x \pm i S_y$ \\
  $\lambda_+=\left(
\begin{array}{ccc}
0& \sqrt{2} & 0 \\
0& 0 & \sqrt{2} \\
0 & 0 & 0
\end{array}
\right)$ , $ \lambda_-=\left(
\begin{array}{ccc}
0& 0 & 0 \\
\sqrt{2}& 0 & 0  \\
0 & \sqrt{2} & 0
\end{array}
\right), $

and $\lambda_z=S_z$.

% The quadratic Hamiltonian for spinor boson in ferromagnetic phase
%
% \begin{widetext}
% \begin{eqnarray}
% &&
% [H_{B}]_{ferro}=-\frac{n_{0}^{2}}{2}(U_{b0}+U_{b2})N+\sum_{{\bf k},\alpha}[\xi_{{\bf k}}+\frac{n_{0}}{2}
% (1-\delta_{\alpha z})(U_{b0}-U_{b2})]\widetilde{a}_{{\bf k}\alpha}^{\dagger}\widetilde{a}_{{\bf k}\alpha}+\frac{n_{0}}{2}(U_{b0}-U_{b2})
% \sum_{{\bf k}}i(\widetilde{a}_{{\bf k}x}^{\dagger}\widetilde{a}_{{\bf k}y}-\widetilde{a}_{{\bf k}y}^{\dagger}\widetilde{a}_
% {{\bf k}x})\nonumber \\
% &&+\frac{n_{0}}{4}(U_{b0}+U_{b2})\sum_{{\bf k}}(\widetilde{a}_{{\bf k}x}\widetilde{a}_{-{\bf k}x}+\widetilde{a}_{{\bf k}x}^{\dagger}
% \widetilde{a}_{-{\bf k}x}^{\dagger}-\widetilde{a}_{{\bf k}y}\widetilde{a}_{-{\bf k}y}-\widetilde{a}_{{\bf k}y}^{\dagger}\widetilde
% {a}_{-{\bf k}y}^{\dagger})
% +\frac{n_{0}}{2}(U_{b0}+U_{b2})\sum_{{\bf k}}i(\widetilde{a}_{{\bf k}x}^{\dagger}\widetilde{a}_{-{\bf k}y}^{\dagger}
% -\widetilde{a}_{{\bf k}x}\widetilde{a}_{-{\bf k}y}).
% \label{a1}
% \end{eqnarray}
% \end{widetext}

% Eq.\ \ref{a1} can then be rearranged as
%
% \begin{eqnarray}
% [H_{B}]_{ferro}=\sum_{{\bf k}>0}\Phi_{{\bf k}\alpha}^{\dagger}
% {g^{0}_{\alpha\beta}}^{-1}\Phi_{{\bf k} \beta}\label{a2}
% \end{eqnarray}
The action $S_B$ when the bosons are in the ferromagnetic phase can
be written using Ref.\ \ref{i2} and \ref{ac1} as
\begin{eqnarray}
 S_B&=&-\frac{1}{\beta}\sum_{\omega_n}\Big[\sum_{{\bf k}}[a^*_{\alpha}({{\bf k},\Omega_n)}
i \Omega_n a_{\alpha}({\bf k},\Omega_n)\nonumber\\
&&-H'_{B}[a^*,a] \Big]. \label{a2}
\end{eqnarray}
$H'_B$ is the quadratic Hamiltonian for spinor boson in ferromagnetic phase and given by,
\begin{widetext}
 \begin{eqnarray}
&&
H'_{B}=-\frac{n_{0}^{2}}{2}(U_{b0}+U_{b2})N+\sum_{k,\alpha}[\xi_{k}+\frac{n_{0}}{2}
(1-\delta_{\alpha
z})(U_{b0}-U_{b2})]a_{k\alpha}^{\dagger}a_{k\alpha}-\frac{n_{0}}{2}(U_{b0}+3U_{b2})\sum_{k}i(a_{kx}^{\dagger}
a_{ky}-a_{ky}^{\dagger}a_{kx})\nonumber \\
&&+\frac{n_{0}}{4}(U_{b0}+U_{b2})\sum_{k}(a_{kx}a_{-kx}+a_{kx}^{\dagger}a_{-kx}^{\dagger}-a_{ky}a_{-ky}-a_{ky}^{\dagger}a_{-ky}^{\dagger})
+\frac{n_{0}}{2}(U_{b0}+U_{b2})\sum_{k}i(a_{kx}^{\dagger}a_{-ky}^{\dagger}-a_{kx}a_{-ky}),
\label{a1}
\end{eqnarray}
\end{widetext}
where , $\xi_{k}=\varepsilon_{k}+4t_{b} =-2t_{b}({\rm
cos}(k_{x}a)+{\rm cos} (k_{y}a)-2)$. We then decouple the boson
fields using Eq.\ (\ref{i8}) and expand about the ferromagnetic
condensate saddle point to obtain the quadratic effective action for
the bosons. This action has the form
\begin{eqnarray}
S_B^{\rm eff} &=& -\frac{1}{\beta}\sum_{\omega_n} \sum_{{\bf k}}
{\mathcal A}^{\ast} ({\bf k},\Omega_n) G_B^{-1} {\mathcal A}({\bf
k},\Omega_n), \label{effac}
\end{eqnarray}
% \begin{eqnarray}
%  \phi_{k\alpha}^{\dagger}=\left(
% %\begin{array}{cc}
% a_{kx},
% a_{-kx}^{\dagger},
% a_{ky},
% a_{-ky}^{\dagger},
% a_{kz},
% a_{-kz}^{\dagger}
% % \end{array}
% \right)^T\label{a3}
% \end{eqnarray}
where ${\mathcal A}({\bf k},\Omega_n)$ denotes the fluctuating boson
fields given by
\begin{eqnarray}
{\mathcal A}^{\ast}({\bf k},\Omega_n) &=& \left[ a^{*}_{x}({\bf
k},\Omega_n), a_{x}(-{\bf k},\Omega_n), a^{*}_{y}({\bf
k},\Omega_n), \right. \nonumber\\
&& \left. a_{y}(-{\bf k},\Omega_n), a^{*}_{z}({\bf k},\Omega_n),
a_{z}(-{\bf k},\Omega_n) \right], \nonumber\\ \label{a3}
\end{eqnarray}
and $G_B^{-1}$ denotes the boson Green's function given by
\begin{eqnarray}
 G_{B}^{-1}=-\left(
\begin{array}{cccccc}
P_{{\bf k}-} & A & -iB & iA & 0& 0\\
A& P_{{\bf k}+} & -iA & iB & 0 & 0\\
iB & iA & P_{{\bf k}-} & -A & 0 & 0\\
-iA & -iB & -A & P_{{\bf k}+} & 0 & 0\\
0 & 0 & 0 & 0 & \xi_{{\bf k}-} & 0\\
0 & 0 & 0 & 0 & 0 & \xi_{{\bf k}+}
\end{array}
\right),\label{a4}
\end{eqnarray}
where, $P_{{\bf k}\pm}=\xi_{{\bf
k}}+\frac{n_{0}}{2}(U_{b0}-U_{b2})\pm i\Omega_n$ , $\xi_{{\bf
k}\pm}=\xi_{{\bf k}}\pm i\Omega_n$,
$A=\frac{n_{0}}{2}(U_{b0}+U_{b2})$ and
$B=\frac{n_{0}}{2}(U_{b0}+3U_{b2})$. Using Eq.\ (\ref{a2}) and Eq.\
(\ref{i3}), we integrate out the bosons by following standard
technique and obtain effective fermionic action as written in Eq.\
(\ref{fham1}) .

%\vspace{5cm}

\end{document}